\documentclass[12pt,preprint]{aastex}

\newcommand{\ha}{H$\alpha$}
\newcommand{\sm}{$\sim$}
\newcommand{\cah}{Ca~{\sc ii}~H}

\usepackage{url}

\begin{document}

\title{STUDY OF TWO SUCCESSIVE THREE-RIBBON SOLAR FLARES ON 2012 JULY 6}

\author{Haimin Wang$^1$, Chang Liu$^1$, Na Deng$^1$, Zhicheng Zeng$^2$, Yan Xu$^1$,  Ju Jing$^1$, and\\ Wenda Cao$^2$}

\affil{1. Space Weather Research Laboratory, New Jersey Institute of Technology, University Heights, Newark, NJ 07102-1982, USA}

\affil{2. Big Bear Solar Observatory, New Jersey Institute of Technology, 40386 North Shore Lane, Big Bear City, CA 92314-9672, USA}

\email{haimin.wang@njit.edu}

\begin{abstract}
This Letter reports two rarely observed three-ribbon flares (M1.9 and C9.2) on 2012 July 6 in NOAA AR 11515, which we found with H$\alpha$ observations of 0$\farcs$1 resolution from the New Solar Telescope and Ca~{\sc ii}~H images from Hinode. The flaring site is characterized with an intriguing ``fish-bone-like'' morphology evidenced by both \ha\ images and a nonlinear force-free field (NLFFF) extrapolation, where two semi-parallel rows of low-lying, sheared loops connect an elongated, parasitic negative field with the sandwiching positive fields. The NLFFF model also shows that the two rows of loops are asymmetric in height and have opposite twists, and are enveloped by large-scale field lines including open fields.  The two flares occurred in succession in half an hour and are located at the two ends of the flaring region. The three ribbons of each flare run parallel to the PIL, with the outer two lying in the positive field and the central one in the negative field. Both flares show surge-like flows in \ha\ apparently toward the remote region, while the C9.2 flare is also accompanied by EUV jets possibly along the open field lines. Interestingly, the 12--25~keV hard X-ray sources of the C9.2 flare first line up with the central ribbon then shift to concentrate on the top of the higher branch of loops. These results are discussed in favor of reconnection along the coronal null-line producing the three flare ribbons and the associated ejections.

\end{abstract}


\section{INTRODUCTION}

During solar flares, two \ha\ ribbons lie in opposite magnetic polarities and run parallel to the magnetic polarity inversion line (PIL). Such a well-known configuration evidences the standard flare model (e.g., Priest \& Forbes 2000), in which the energy release along field lines generates flare emissions at the two footpoints in the lower atmosphere. Groups of footpoints along a series of coronal loops form two ribbons, and they move away from the PIL as successive reconnections proceeds into the higher corona (e.g., Wang et al. 2003). In another widely known tether-cutting reconnection model (Moore et al. 2001; Liu et al. 2013), the interaction of two sigmoidal magnetic structure produces four footpoint sources, which later evolve to two ribbons following the re-closing of the opened envelope fields (e.g., Liu et al. 2007a). In some rare cases, two ribbon-like emissions were also observed in hard X-rays (HXRs) (Liu et al. 2007b). Although flares with three parallel ribbons in \ha\ were reported in some early studies (Ogir \& Antalova 1986; Antalova \& Ogir 1988), the central ribbon was found as signatures of sources at the top, rather than the feet, of flaring loops.

With high-resolution observations, flares with a closed circular-like ribbon have also been revealed (e.g., Masson et al. 2009; Reid et al. 2012; Wang \& Liu 2012, hereafter WL12). The associated magnetic field configuration usually consists of a parasitic region encompassed by the opposite-polarity field, forming a circular PIL. Besides the circular ribbon, an inner and an outer (remote) ribbons are often found. Some flares also exhibit surge-like eruptions from the area of the circular ribbon. Recently, the hot coronal loops connecting a quasi-circular chromospheric ribbon and a remote brightening were analyzed by Sun et al. (2013) based on observations from the Solar Dynamics Observatory (SDO; Pesnell et al. 2012). In interpreting all these events, the fan-spine magnetic topology is involved, where the dome-shaped fan portrays the closed separatrix surface and the inner and outer spine field lines in different connectivity domains pass through a coronal null point (Lau \& Finn 1990; T{\"o}r{\"o}k et al. 2009). Interestingly, the outer spine can be closed and/or open (see Figure 1 of WL12). In the former case, the slipping/slip-running reconnection within the quasi-separatrix layers (QSLs) that embed the fan and spine separatrices leads to the sequential brightening of the circular fan ribbon, while the ensued null-point reconnection causes the later remote brightening (Masson et al. 2009). In the latter case, the null-point reconnection generates surges/jets that erupt outward (Pariat et al. 2009, 2010).

The state-of-the-art observation with high spatiotemporal resolution of the 1.6~m New Solar Telescope (NST; Goode et al. 2010; Cao et al. 2010) at Big Bear Solar Observatory (BBSO) allows an assessment of the low atmospheric structure in an unprecedented detail. In this Letter, we report two closely spaced three-ribbon flares with GOES-class M1.9 and C9.2 in the flare-productive NOAA AR 11515 on 2012 July 6. The properties of the three flare ribbons and the associated flaring loops and surges are conspicuously observed in NST \ha\ images. We will discuss the formation of the multiple ribbons and the associated ejections in the context of the fan-spine magnetic topology, with the help of a nonlinear force-free field (NLFFF) extrapolation from the active region and other multiwavelength observations.

\section{OBSERVATIONS AND DATA PROCESSING}

The 2012 July 6 M1.9 and C9.2 flares started at 18:48 and 19:20~UT, peaked at 18:55 and 19:24~UT, and ended at 19:05 and 19:26~UT, respectively. With a 76-element adaptive optics system and speckle-masking image reconstruction using 100 frames, BBSO/NST achieved diffraction-limited imaging of both flares in \ha\ line center, H$\alpha$ blue wing ($\Delta\lambda=-$~0.75~\AA; lower-middle chromosphere), He~{\sc i} 10830~\AA\ blue wing ($\Delta\lambda=-$~0.2~\AA; high corona), and TiO band (a proxy for the continuum photosphere at 7057~\AA). The images have a spatial resolution of 0$\farcs$1--0$\farcs$17 and a cadence ranging from 6 to 15~s. This flaring active region was also the target of Hinode with limited \cah\ and spectro-polarimetric observations (Tsuneta et al. 2008).

The full-disk vector magnetic field data with a 12 minute cadence from the Helioseismic and Magnetic Imager (HMI; Schou et al. 2012) on board SDO are derived using the VFISV inversion code by Borrero et al. (2011). The 180$^{\circ}$ azimuthal ambiguity is resolved with the minimum energy method (Metcalf 1994; Leka et al. 2009). We chose the latest version in Space weather HMI Active Region Patches (SHARP) (Turmon et al. 2010) with a remapped format using Lambert equal area projection. The observed fields have been transformed to heliographic coordinates (Gary \& Hagyard 1990). The SHARP data we used is at a preflare time (18:36~UT) and covers the entire AR 11515 as well as the nearby AR 11514 with balanced magnetic fluxes. We caution that the quality of vector magnetograms may be affected as this region is close to the western limb (with an orientation cosine factor about 0.6)(Rudenko \& Anfinogentov 2011). After preprocessing the photospheric boundary to best suit the force-free condition (Wiegelmann et al. 2006), we constructed NLFFF models using the ``weighted optimization'' method (Wiegelmann 2004) with the error treatment incorporated (Wiegelmann \& Inhester 2010; Wiegelmann et al. 2012). The calculation was performed using the 2~$\times$~2 rebinned magnetogram within a box of 576~$\times$~232~$\times$~256 uniform grid points, which corresponds to 420~$\times$~169~$\times$~187~Mm$^3$. In addition, the full-Sun potential field source surface (PFSS; Schrivjer \& De Rosa 2003) model available in the Solar Software was resorted to for exploration of characteristic magnetic topology, although it lacks the resolution to deal with small-scale structures.

To further examine the coronal signatures of flares and ejections, we used images in multiple bands taken by the Atmospheric Imaging Assembly (AIA; Lemen et al. 2012) on board SDO. An AIA 4500~\AA\ (white light) image was also utilized to register NST and Hinode images by matching sunspot and plage areas, with an alignment accuracy of about 1\arcsec. Furthermore, flare HXR emission was recorded by the Reuven Ramaty High Energy Solar Spectroscopic Imager (RHESSI; Lin et al. 2002). RHESSI PIXON images (Hurford et al. 2002) in the energy range of 12--25~keV were reconstructed using detectors 1--8 with a 24~s integration time.

\section{RESULTS}
We first investigate the magnetic field structure of the flaring region, and then concentrate on describing major observational features in \ha, 10830~\AA, HXR, and EUVs. More flare dynamics can be seen in the accompanying animations in the online journal associated with Figures 3 and 4.

\subsection{Fish-bone-like Magnetic Structure}
The flaring region is characterized with two $\delta$ spots p1-n and p2-n (Figures~\ref{f1}(a) and (b)), where an elongated strip n of negative magnetic polarity is shared by positive fluxes p1 and p2 with a shape of magnetic tongues (e.g., Luoni et al. 2011). The penumbral filaments lying between n and p1/p2 and also the Hinode magnetic vectors clearly indicate highly sheared magnetic fields along the PIL (the white line in Figure~\ref{f1}(b)). It is noticeable that the PIL makes a closed turn in the southwest but is ``open'' in the northeast, as the negative field stretches to a remote region S (also see Figure~\ref{f2}(d)). The overall morphology of magnetic field, although much extended and elongated, is analogous to the circular flare regions for the parasitic configuration.

In order to delineate the characteristic structure of the flare volume, in Figures~\ref{f2}(a) and (b), we trace magnetic fields from the minor negative polarity region n in our NLFFF model at a preflare time 18:36~UT, and separate the field lines landing at the positive p1 and p2 regions by using red and blue colors, respectively. The color depth also denotes the loop height. The extrapolation result evidently shows that the sheared fields stemming from n consist of closed field lines at chromospheric heights (\sm2.1~Mm), and that these field lines bifurcate cleanly into two semi-parallel rows of loops, forming a ``fish-bone''-like structure. Interestingly, the northern (red) and southern (blue) branches of loops show a distinct asymmetry in height, with the mean value of the former about double that of the latter. Another difference between them is found in terms of magnetic twist. We compute the twist index $T_n$ (in number of turns) defined by Inoue et al. (2011) as $T_n = \frac{1}{4\pi} \overline{\alpha} L$, where $\overline{\alpha}$ is the mean force-free parameter derived for the two feet of a field line on the photosphere, and $L$ is the loop length (Liu et al. 2013). It can be obviously seen in Figure~\ref{f2}(c) that nearly all the northern (southern) branch of loops possess a negative (positive) twist with $\overline{T_n}=-0.34$ (0.31). As to connectivity in a large scale, it is apparent that the above fish-bone structure is completely embedded under the overlying loops connecting from p1/p2 to the negative field region S in the northeast (Figure~\ref{f2}(d)). Moreover, both the PFSS and NLFFF models exhibit open field lines in this flaring region (cyan lines in Figures~\ref{f1}(b) and \ref{f2}(a)), which originate from the western portion of the positive flux.

\subsection{Three Flare Ribbons and the Associated Surges/Jets}

\begin{figure}
\epsscale{1.01}
\plotone{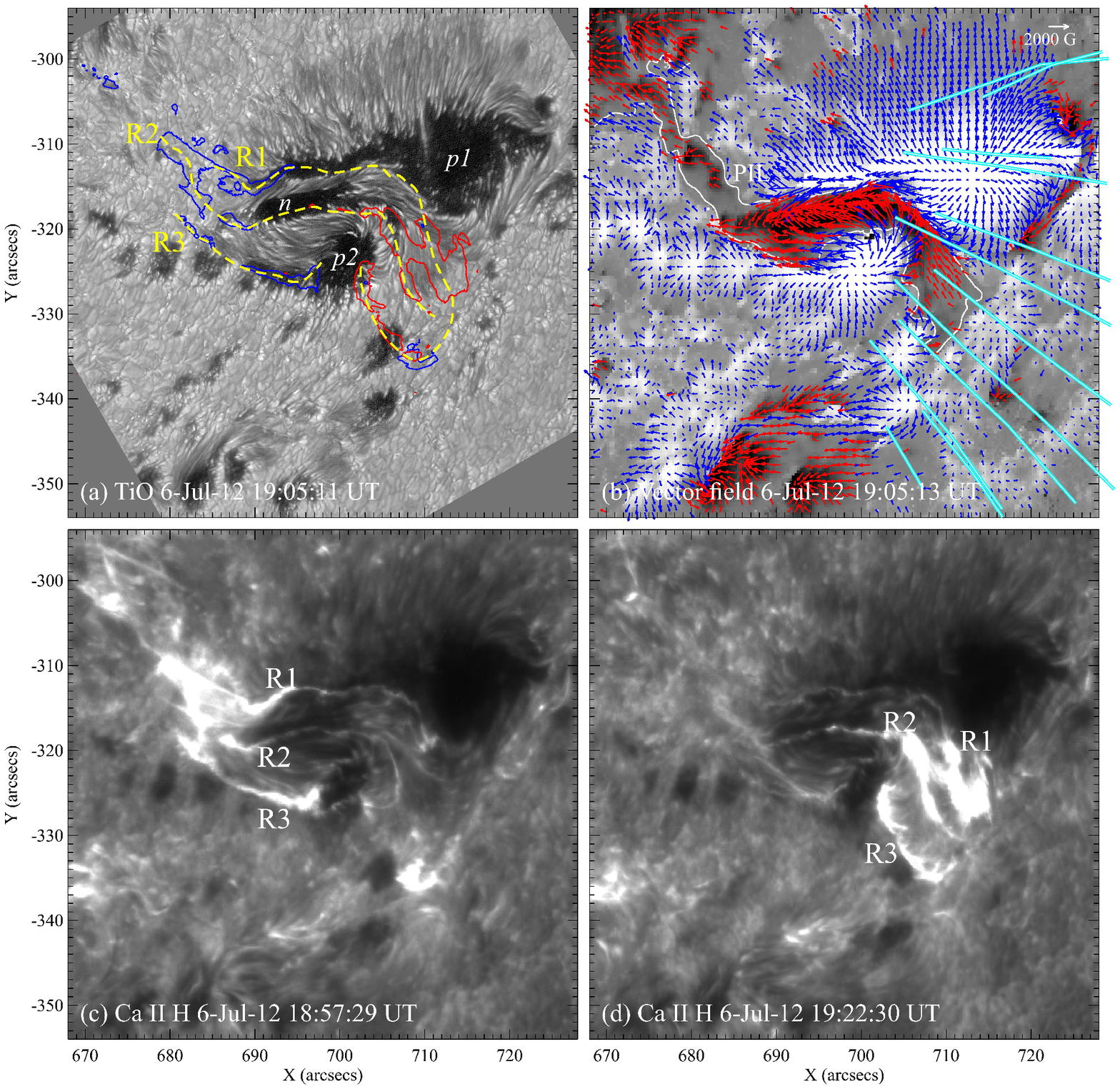}
\caption{NST TiO (a), Hinode vector magnetic field (b), and Hinode Ca~{\sc ii}~H (c--d) images near the peak times of the 2012 July 6 M1.3 (18:55~UT) and C9.2 (19:24~UT) flares. In (a), the blue and red contours outline the flare ribbons in (c) and (d), and the dashed lines illustrate the three ribbons R1--R3. The cyan lines in (b) represent open fields in the PFSS model with 1$^{\circ}$ resolution, and the white line is the flaring PIL. All the images in this paper are aligned with respect to 2012 July 6 19:24:26~UT. \label{f1}}
\end{figure}

\begin{figure}
\epsscale{1}
\plotone{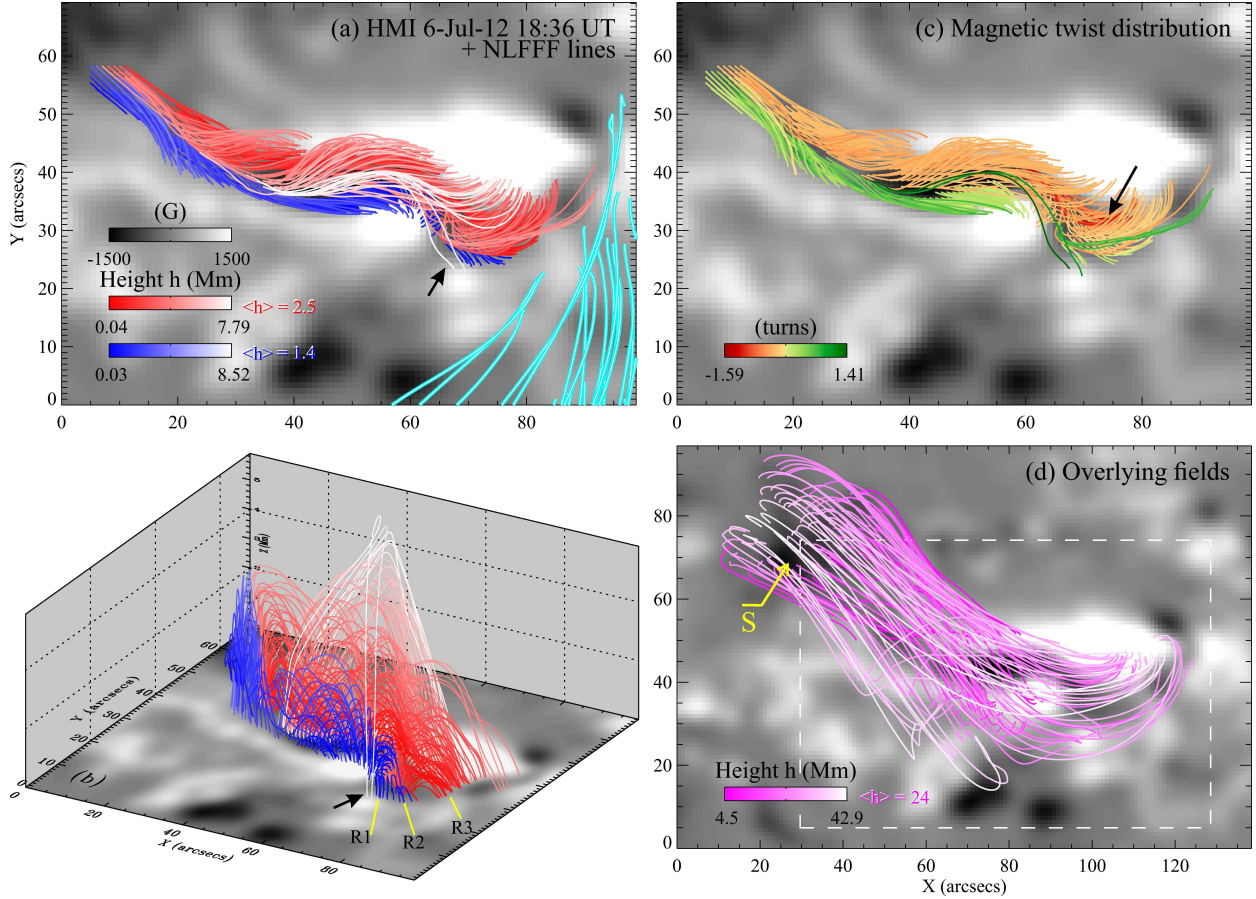}
\caption{A remapped and preprocessed HMI vertical field in top (a, c, and d) and perspective (b) views, overplotted with NLFFF lines traced from the parasitic negative field (with $|B| > 200$~G) of the flaring region (a, b, and c) and a remote negative field region S at northeast (d). In (a) and (b), the field lines land at the northern and southern positive field regions are colored red and blue, respectively. The cyan lines in (a) are open fields. The field lines in (c) are colored according to the magnetic twist index. About 18\% of field lines in (a) do not appear in (b) as their two feet have $\alpha$ with opposite sign. The dashed box in (d) represents the plotted region of (a) and (c) as well as the bottom boundary of (b). \label{f2}}
\end{figure}

The modeled magnetic structure can be further understood by studying the properties of flare ribbons in the low atmosphere and vice versa. In Figures~\ref{f1}(c) and (d), we present Hinode Ca~{\sc ii}~H images closest to the maximum time of the M1.3 and C9.2 flares, both of which show three ribbons located at either end of the fish-bone-like region. Intriguingly, these ribbons seem to trace three bright channels in \cah\ (illustrated as dashed lines in Figure~\ref{f1}(a)), and we label them R1--R3 in both flares. By comparing with TiO and magnetic field images (Figures~\ref{f1}(a) and (b)), it is shown that the central ribbon R2 and the outer ribbons R1/R3 are in the negative (n) and positive polarity (p1 and p2) regions, respectively. In the C9.2 flare, R1 and R3 also seem to join in the southwest (see further the \ha\ images discussed below), which is consistent with the ``closed'' PIL there (cf. Figures~\ref{f1}(a) and (b)). It is evident that the present three-ribbon flares are essentially different from those studied by Ogir \& Antalova (1986) and  Antalova \& Ogir (1988), in which the third central ribbon is located above PILs. Since the \cah\ filter of Hinode/SOT has an average response height of \sm0.25~Mm (Carlsson et al. 2007), it is highly likely that all the three ribbons originate from a low altitude.

\begin{figure}
\epsscale{1.0}
\plotone{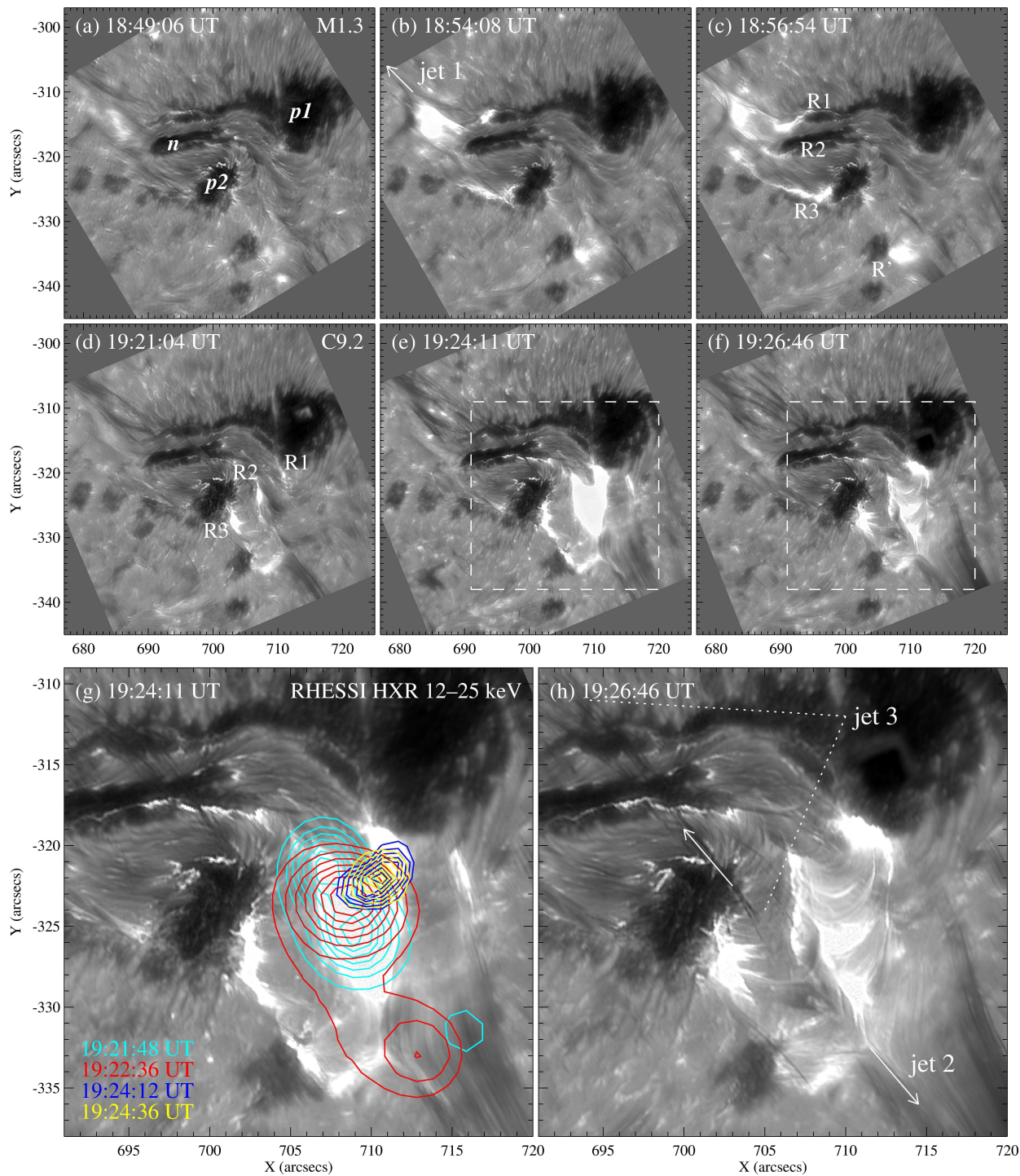}
\caption{Time sequence of NST H$\alpha - 0.75$~\AA\ images showing the evolution of the M1.3 (a--c) and C9.2 (d--f) flares. The dash-boxed region in (e) and (f) is magnified in (g) and (h), respectively. Contours (30\%, 40\%, 50\%, 60\%, 70\%, 80\%, and 90\% of each maximum flux) in (g) represent RHESSI 12--25~keV PIXON images at the event onset and HXR peak. \label{f3}}
\end{figure}

The ribbons R1--R3 are also present in NST \ha~$-$~0.75~\AA\ images, which better display the event evolution with a high cadence. It can be seen from Figures~\ref{f3}(a)--(c) and the time-lapse movie that during the M1.3 flare, R1 propagates from west to east, while R3 extend southwestward in general. Although R3 stops at the negative spot n, the region R$^{\prime}$ (see Figure~\ref{f3}(c)) is also brightened, possibly due to the connectivity between the flaring region and this location (see the long loops pointed to by the arrow in Figures~\ref{f2}(a) and (b)). Hence the outer ribbons of this flare undergo a seemingly sequential brightening in the counterclockwise direction, similar to that reported in circular ribbon flares (e.g., WL12). Another prominent activity is the \ha\ surge (the arrow in Figure~\ref{f3}(b)), which envelopes the three ribbons and converges toward the S region in the northeast presumably along large-scale loops. The surge-like flows are discernible in \ha\ before the flare, but during which an abrupt enhancement is obvious. The large-scale loops are also brightened in EUVs (see the accompanied animations), similar to those described by Sun et al. (2013) in a circular-ribbon event.

The C9.2 flare occurred about half an hour later illuminates the flaring structure more clearly (see Figures~\ref{f3}(d)--(h)). We observe that (1) ribbons R1 and R3 show a northward propagation at a speed of \sm20~km~s$^{-1}$, which is comparable to that of the previously reported moving flare sources along PILs (e.g., Liu et al. 2010). Similar to those in \cah\ images, R1 and R3 also seem to join at the southern end. (2) Two rows of semi-parallel arcade loops connect R1/R2 and R2/R3, which well agrees with our NLFFF model (also see an \ha\ center-line image in Figure~\ref{f4}(b)). The loops R2--R1 are brighter in \ha\ and some EUV bands (e.g., the 335~\AA\ image in Figure~\ref{f4}(c)), and may correspond to the group of more twisted ($T_n \approx$1) loops as pointed to by an arrow in Figure~\ref{f2}(c). The bright ribbon R1 is also cospatial with the source of sunquakes in this event (S. Zharkov et al. 2013, private communication). (3) At the event onset, there are extended 12--25 keV HXR sources that line up with the ribbon R2; later at the HXR peak, the more compact HXR emission is most probably at the top of the loops R2--R1 (Figure~\ref{f3}(g)) (with some ambiguity due to the projection effect). The centroid of HXR sources thus shifts northwestward, the motion of which is consistent with that of the brightenings in hot AIA channels (e.g., 94~\AA; see the animation). (4) A chain of surges emerge from multiple cusp-shaped locations above R2, and a fountain-like spray also envelopes the loops R2--R1. Notably, there is a relatively thin thread of surge (labeled jet 3 in Figure~\ref{f3}(h)) that shoots northwestward, which is in a nearly perfectly opposite direction of the main surge (labeled jet 2). The jet 2 can be observed in \ha\ center line, 10830~\AA, and EUVs, while the jet 3 is missing from 10830~\AA\ and cannot be distinguished from the brightened hot EUV loops connecting the flare to the remote region S (see Figure~\ref{f4}). We surmise that the jet 3 in \ha\ moves along closed loops reflecting flows of cooler materials. In contrast, the hot jet 2 ejects outward along open field lines as modeled in this area. It is also worth mentioning that the jets 2 and 3 show sustained activity from \sm19:11~UT even before the C9.2 flare, which may imply a continuous mild energy release.

\section{SUMMARY AND DISCUSSION}

We have presented a detailed study of the M1.3 and C9.2 flares on 2012 July 6 taking advantage of both the unprecedented high-resolution NST observations and NLFFF modeling. We report, for the first time, that the three parallel ribbons along the PIL in these flares are all in the feet of flaring loops. The main results are summarized as follows.

\begin{enumerate}

\item The flaring region is featured with an elongated, parasitic negative field on the surface and a fish-bone-like magnetic structure in 3D, which consists of two semi-parallel rows of closed loops with highly asymmetric heights at chromospheric altitude. Large-scale envelope fields connect the flaring site to a remote region S, and open fields are present near the southwestern end of the region.

\item Each flare exhibits three ribbons R1--R3 clearly observed in both \cah\ and \ha, and the flares are both accompanied with surges. The central and outer ribbons lie in negative and positive magnetic field regions, respectively, and the latters show obvious motion along the PIL. In the C9.2 flare, two rows of bright loops R3--R2 and R2--R1 are conspicuously seen, and a series of surges/jets originate from the cusp-shaped locations lining above the central ribbon R2. Moreover, HXR sources first spread along R2 at the flare onset, then shift to concentrate on the top of higher loops R2--R1.

\end{enumerate}

The peculiar magnetic structure, as indicated by our NLFFF model and high-resolution \ha\ images, and the unusual three flare ribbons along the PIL pose a challenge for understanding the flare mechanism. Nevertheless, we note that the morphology of this parasitic region and its topological components are strongly reminiscent of those characterize the circular flare active region (WL12). We speculate that this flaring region, in general, consists of the same cross section as that of the fan-spine magnetic topology (e.g., Pariat et al. 2010; WL12), which repeats in the third dimension (meaning a 2.5D symmetry) to produce a fish-bone-like structure. Accordingly, the single isolated coronal null-point in the 2D fan-spine fields would be prolonged to form a null line. This conjecture is schematically illustrated in Figure~\ref{f5}, where a semi-circular fan-spine element is also incorporated. Such fan-spine field lines in a translational symmetry was modeled before using MHD simulations (e.g., Edmondson et al. 2010, and references therein), where reconnection can be induced along the null-line when the bottom photosphere is subject to perturbations from horizontal flows. Nevertheless, further topological analysis in a separate investigation is worthwhile to pinpoint null points (e.g., Sun et al. 2013) and substantiate the envisioned null line.

\begin{figure}
\epsscale{1.}
\plotone{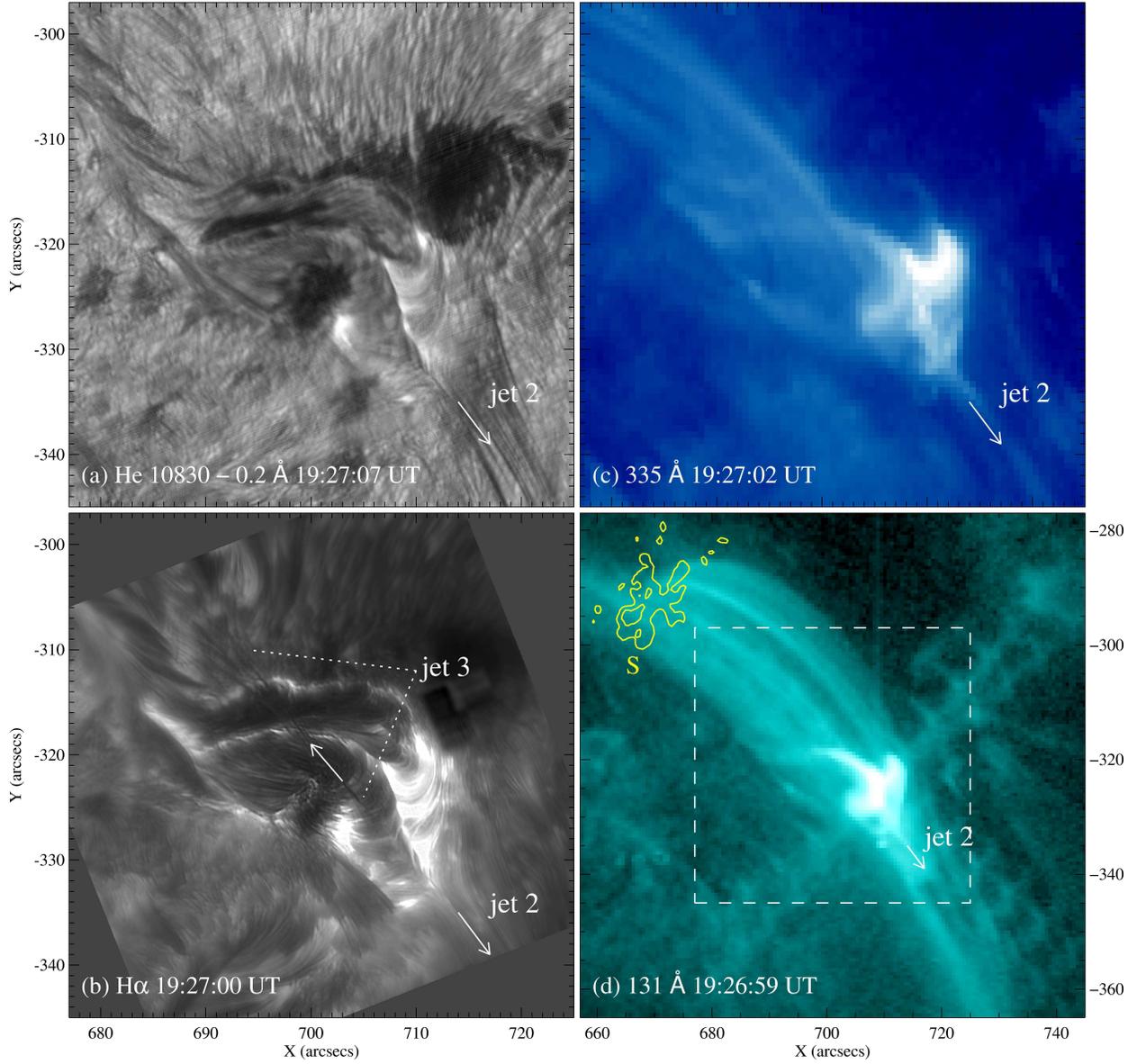}
\caption{NST He~{\sc i} 10830~$-$~0.2~\AA\ (a), H$\alpha$ line center (b), SDO/AIA 335~\AA\ (c), and 131~\AA\ (d) images right after the C9.2 flare. In (d), LOS magnetic fields at $-$700 and $-$300~G of the remote region S are overplotted as yellow contours. The dashed box is the FOV of (a)--(c). \label{f4}}
\end{figure}

\begin{figure}
\epsscale{1}
\plotone{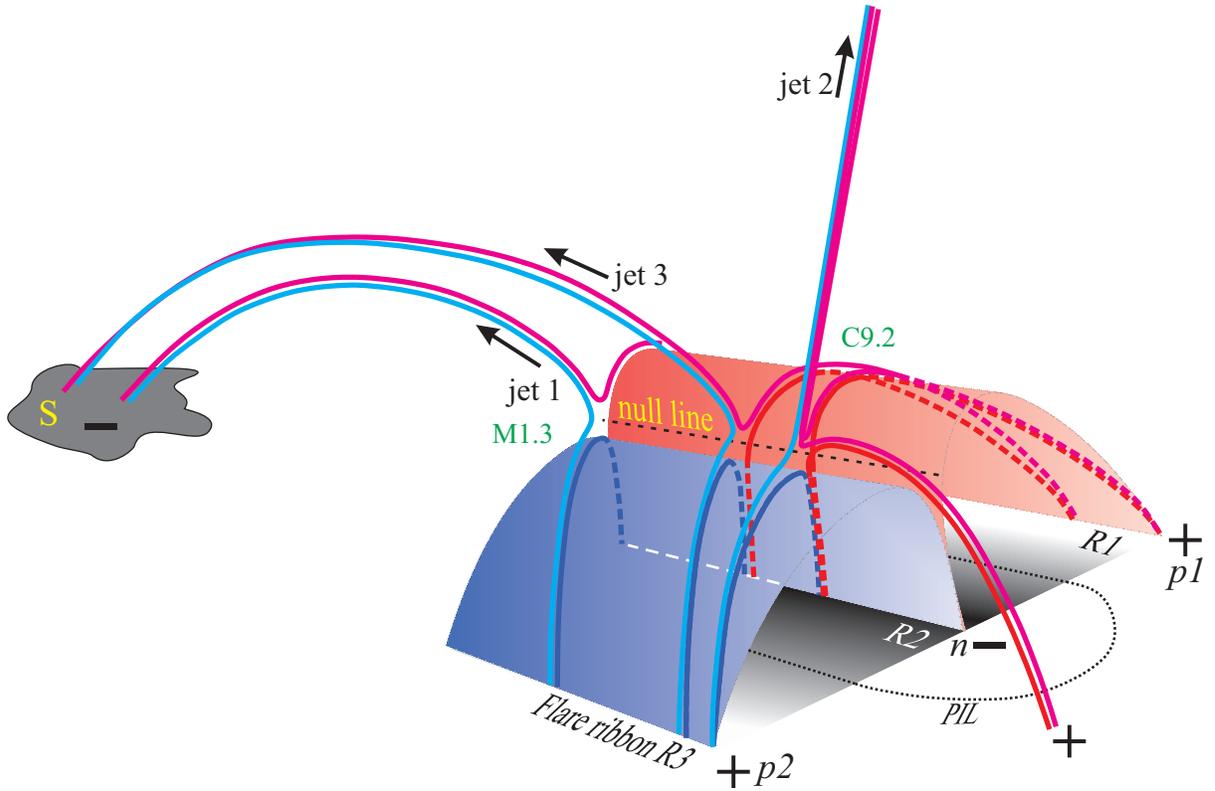}
\caption{Schematic picture demonstrating the relationship between three flare ribbons R1--R3 and jets 1--3 in a 3D null-line magnetic structure, based on our NST observation, the NLFFF extrapolation, and a previous picture of WL12. See Section~4 for details. \label{f5}}
\end{figure}

We argue that many features of the above flares of interest can be accommodated in this scheme with additional plausible modifications (see Figure~\ref{f5}). First, the triple flare ribbons R1--R3 could map the photospheric intersections of the fan and spine field lines, which stem from the null line. The associated chain of surges can naturally form following the reconnection along the null line. The characteristic sequential brightening signifying QSL reconnections (Masson et al. 2009) is also observed in the M1.3 flare. Second, the outer spine fields can return to a remote region (in this case, region S) or open upward (WL12). The modeled large-scale connection between the flaring and S regions is also clearly visible in coronal images (Figure~\ref{f4}(d)). Third, the fan loops R3--R2 and R2--R1 are highly asymmetric in height. This condition with the most probably tilted outer spine would result in strong, extended current sheets, which have been demonstrated highly favorable for reconnection (Pariat et al. 2009, 2010; Edmondson et al. 2010). We suggest that our observed northwestward shifting of HXR and EUV sources from above the central R2 toward the tops of higher loops R2--R1 might be a manifestation of the reconnection along the induced current sheet. Finally, as portrayed based on the extrapolation result, the loops R3--R2 and R2--R1 are oppositely twisted and both sheared with respect to the PIL.

In summary, the occurrence of the three flare ribbons, the two rows of \ha\ loops and the accompanied surges/jets, and the NLFFF model result lead us to speculate on a fan-spine field topology in a 2.5D symmetry, where a null line is embedded. Further studies on the formation and topology of the fish-bone-like structure may shed insights on the dynamics of 3D magnetic reconnection.

\acknowledgments

We thank the BBSO staff for the NST observations, the SDO and Hinode teams for the magnetic and flare images, T. Wiegelmann for the NLFFF extrapolation code, and the referee for valuable comments. This work was supported by NSF AGS grants 1153226, 1153424, 1250374, 0847126, and 0745744, and by NASA grants NNX11AO70G, NNX11AC05G, NNX11AQ55G, NNX13AG13G, NNX13AF76G, NNX14AC12G, and\\ NNX08BA22G.

\clearpage

\end{document}